\documentstyle[12pt]{article}
\newlength{\extraspace}
\newlength{\extraspaces}
\begin{document}

\thispagestyle{empty}

\hfill \parbox{3.5cm}{hep-th/ \\ SIT-LP-01/10}
\vspace*{1cm}
\begin{center}
{\large \bf New Einstein-Hilbert Type Action for Unity of Nature} \footnote{Talk given at 100 Years Werner Heisenberg 
-Works and Impact-, September 26-30, 2001, Bamberg, Germany. 
To appear in the Proceeding: {\it Fortschritte der Physik}(Springer-Verlag, 2002). 
 \\ e-mail:shima@sit.ac.jp }\\[20mm]
{\bf Kazunari SHIMA} \\[2mm]
{\em Laboratory of Physics,  Saitama Institute of Technology} \\
{\em Okabe-machi, Saitama 369-0293, Japan}\\[2mm]
{November  2001}\\[15mm]


\begin{abstract}
A new Einstein-Hilbert(E-H) type (SGM) action is obtained by performing the 
Einstein gravity analogue geomtrical arguments in high symmetric (SGM) spacetime. 
All elementary particles except graviton are regarded as the eigenstates of SO(10) super-Poincar\'e 
algebra(SPA) and composed of the fundamental fermion "superons" of nonlinear supersymmetry(NL SUSY). 
Some phenomenological implications and the linearlization of the action are discussed briefly.    \\

\end{abstract}
\end{center}

\newpage

\section{Introduction}
The standard model(SM) is established  
as a unified model for the electroewak interaction. 
Nevertheless, it is very unsatisfactory 
in many aspects, e.g. it can not explain the particle quantum numbers $(Q_{e},I,Y,color, i.e. 
1 \times 2 \times 3 gauge \ structure)$, the three-generations structure and contains more than 28 
arbitrary parameters(in the case of neutrino oscillations) even disregarding the mass 
generation mechanism for neutrino. The simple and beautiful extension to SU(5) GUT has serious 
difficulties, e.g. the life time of proton, etc and is excluded so far. 
The SM and GUT equiptted naively with supersymmetry(SUSY) 
have improved the situations, e.g. the unification 
of the gauge couplings at about $10^{17}$, relatively stable proton(now threatened by experiments),etc., 
but they posess more than 100 arbitrary parameters and less predictive powers. 
However SUSY\cite{WB} is an essential notion to unify various topological and non-topological charges and 
gives a natural framework to unify spacetime and matter leading to the birth of supergravity(SUGRA).  
Unfortunately the maximally extended SO(8) SUGRA is too small to accommodate all observed particles 
as elementary fields. The straightforward extension to SO(N) SUGRA with ${N>9}$ has a difficulty 
due to so called the no-go theorem on the massless elementary high spin$(>2)$ (gauge) field. 
The massive high-spin is another.   \\
Furthermore, we think that from the viewpoint of simplicity and beauty of nature 
it is interesting to attempt the accommodation of all observed  particles in a single
irreducible representation of a certain algebra(group) especially in the case of 
high symmetric spacetime having a certain boundary(,i.e. a boundary condition)  
and the dynamics  are 
described by the spontaneous breakdown of the high symmetry of spacetime by itself, which is encoded 
in the nonliner realization of the geometrical arguments of  spacetime. 
Also the no-go theorem does not exclude the possibility that the fundamental action, if it exists,
posesses the high-spin degrees of freedom not as the elementary fields but as some ${\em composite}$
eigenstates of a certain symmetry (algebra) of the fundamental action. 
In this talk we would like to present a model along this scenario.       \par
\section{Superon-Graviton Model(SGM)-Phenomenolgy-}   
Among all single irreducible representations of all SO(N) extended super-Poincar\'e(SP) symmetries, 
the massless irreducible representations of SO(10) SP algebra(SPA) is the only one that accommodates 
minimally all observed particles including the graviton\cite{ks1}. 
10 generators  $Q^{N}(N=1,2,..,10)$ of SO(10) SPA are the
fundamental represemtations of SO(10) internal symmetry  and decomposed  
$\underline{10} =  \underline 5+ \underline 5^{*}$  with respect to SU(5) following 
$SO(10) \supset SU(5) $.
For the  massless  case the little algebra of SO(10) SPA for the
supercharges in  the light-cone frame $P_{\mu}=\epsilon(1,0,0,1)$  becomes
after a suitable rescaling
\begin{equation}
\{ Q_{\alpha}^{M}, Q_{\beta}^{N} \}
=\{ \bar{Q}_{\dot\alpha}^{M}, \bar{Q}_{\dot\beta}^{N} \}=0, \quad
\{Q_{\alpha}^{M},\bar{Q}_{\dot\beta}^{N}\}
={\delta}_{{\alpha}1}{\delta}_{{\dot\beta}{\dot1}}{\delta}^{MN},
\label{algebra}
\end{equation}
where $\alpha,\beta=1,2$ and $M,N=1,2,...5$. 
By identifying the graviton with the Clifford vacuum $\mid\Omega\rangle$
(SO(10) singlet) satisfying
$Q_{\alpha}^{M} \mid\Omega\rangle=0$. 
and performing the ordinary procedures we obtain  $2\cdot2^{10}$ dimensional irreducible representation 
of the little algebra (\ref{algebra}) of SO(10) SPA as follows\cite{ks1}:     
$\Bigl[\underline{1}(+2), \underline{10}(+{3 \over 2}),
\underline{45}(+1), \underline{120}(+{1 \over 2}),
\underline{210}(0),
\underline{252}(-{1 \over 2}),  \\
\underline{210}(-1),
\underline{120}(-{3 \over 2}),
\underline{45}(-2), 
\underline{10}(-{5 \over 2}), \underline{1}(-3)\Bigr]
+ \Bigl[ \mbox{CPT-conjugate} \Bigr]$,                          \\
where $\underline{d}(\lambda)$  represents
SO(10) dimension $\underline{d}$ and the helicity $\lambda$.                   
By noting that the helicities of these  states are
automatically determined by SO(10) SPA in the light-cone and that
$Q_{1}^{M}$ and  $\bar{Q}_{\dot{1}}^{M}$
satisfy the algebra of the annihilation and the creation operators for the
massless spin ${1 \over 2}$ particle, we speculate boldly that
these massless states spanned upon the  Clifford vacuum  $\mid\Omega(\pm2)\rangle$
are the massless (gravitational) eigenstates of spacetime and matter 
with SO(10) SP symmetric structure, which are composed of the
fundamental massless object $Q^{N}$, $\it{superon}$ with spin ${1 \over 2}$.
Because they correspond merely to all possible nontrivial combinations of the multiplications of
the spinor charges(i.e. generators) of SO(10) SP algebra(clustering by a universal force?).
Therefore we regard $\underline 5+ \underline 5^{*}$  as $\it{a}$ ${superon}$-${quintet}$
and $\it{an}$ ${antisuperon}$-${quintet}$. The speculation is dicussed later.
To survey the physical implications of superon model for matter we assign tentatively the 
following SM quantum numbers to superons and adopt the following symbols. 
\begin{eqnarray}\underline{10} & = & \underline 5
+ \underline 5^{*}  \nonumber \\
& = & \Bigr[ Q_{a}(a=1,2,3) ,Q_{m}(m=4,5)\Bigl] + \Bigr[Q_{a}^{*}(a=1,2,3),Q_{m}^{*}(m=4,5)\Bigl], \nonumber  \\
& = & [(\underline 3, \underline 1;-{1 \over 3},-{1 \over 3},-{1 \over 3}), 
 (\underline 1, \underline 2;1, 0)] +
[(\underline 3^{*}, \underline 1;{1 \over 3},{1 \over 3},{1 \over 3}), 
 (\underline 1, \underline 2^{*};-1,0)],
\end{eqnarray}
where we have specified (${\underline{SU(3)},\underline{SU(2)}}$; electric  charges ) and 
${a=1,2,3}$ and ${m=4,5}$ represent the color and electroweak components of
superons respectively. 
Interestingly our model needs only five superons  which have 
the same quantum numbers  as the fundamental matter multiplet ${\underline 5}$ of SU(5) GUT 
and satisfy the Gell-Mann--Nishijima relation.     \\
\begin{equation}
Q_{e}=I_{z} + {1 \over 2}(B-L).
\end{equation}
Accordingly all ${2 \cdot 2^{10}}$  states are specified uniquely with respect to ( $SU(3), SU(2)$; electric charges ).
Here we suppose drastically that a field  theory of SGM exists and all unnecessary 
(for SM) higher helicity states become massive in $SU(3) \times SU(2) \times U(1)$ invariant way 
by eating the lower helicity states corresponding to the superHiggs mechanism 
and/or to the diagonlizations of the mass terms of the high-spin fields via 
[SO(10) SPA upon the Clifford vacuum]
$\rightarrow$ [ \ $SU(3) \times SU(2) \times U(1)$ ] 
$\rightarrow$ [ \ $SU(3) \times U(1)$ ]. 
We have carried out the recombinations of the states and 
found surprisingly that all the massless states necessary  for 
the SM with three generations of quarks and leotons appear in the surviving 
massless states (therefore, no sterile neutrinos). 
Among predicted new particles 
one lepton-type electroweak-doublet $( \nu_{\Gamma}, \Gamma^{-} )$ with spin ${3 \over 2}$ 
with the mass of the electroweak scale $( \leq Tev)$ 
and  doublly charged heavy$( >Tev)$ leptons are color singlets and can be observed directly.  \\
As for the assignments of observed particles, we take for simplicity   
the following left-right symmetric assignment for quarks and  leptons 
by using the conjugate representations  naively , i.e. 
$( \nu_{l}, \it{l}^{-} )_{R}= (\bar\nu_{l}, \it{l}^{+} )_{L}$, etc\cite{ks2}. 
Furthermore as for the generation assignments we assume simply that the states 
with more (color-) superons turn to acquiring larger masses in the low energy and 
no a priori mixings among genarations.
The surviving massless states identified with  SM(GUT) are  as follows.                       \\
For three generations of leptons
[$({\nu}_{e}, e)$,  $({\nu}_{\mu}, \mu)$,  $({\nu}_{\tau}, \tau)$], we take
\begin{equation}
\Bigr[(Q_{m}{\varepsilon}_{ln}Q_{l}^{*}Q_{n}^{*}),
(Q_{m}{\varepsilon}_{ln}Q_{l}^{*}Q_{n}^{*}Q_{a}Q_{a}^{*}),
(Q_{a}Q_{a}^{*}Q_{b}Q_{b}^{*}Q_{m}^{*})\Bigl]
\end{equation}
\noindent and the conjugate states respectively.
\\
For three generations of quarks [$( u, d )$, $( c, s )$, $( t, b )$],
we have ${\em uniquely}$
\begin{equation}
\Bigr[({\varepsilon}_{abc}Q_{b}^{*}Q_{c}^{*}Q_{m}^{*}),
({\varepsilon}_{abc}Q_{b}^{*}Q_{c}^{*}Q_{l}
{\varepsilon}_{mn}Q_{m}^{*}Q_{n}^{*}),
({\varepsilon}_{abc}Q_{a}^{*}Q_{b}^{*}Q_{c}^{*}Q_{d}Q_{m}^{*})\Bigl]
\end{equation}
\noindent and the conjugate states respectively. 
For $SU(2) \times U(1)$ gauge bosons [ ${{W}^{+},\ Z,\ \gamma,\ {W}^{-}}$], 
$SU(3)$ color-octet gluons [${G^{a}(a=1,2,..,8)}$], 
[$SU(2)$ Higgs Boson], [$( X,Y )$] leptoquark bosons in GUTs, 
and  a color- and $SU(2)$-singlet neutral gauge boson from
${{\underline 3} \times {\underline 3^{*}}}$ (which we call simply $S$  boson
to represent the singlet) we have   \\
${[Q_{4}Q_{5}^{*}, {1 \over \sqrt{2}}( Q_{4}Q_{4}^{*} \pm Q_{5}Q_{5}^{*}),
Q_{5}Q_{4}^{*}]}$, \\
${[Q_{1}Q_{3}^{*},Q_{2}Q_{3}^{*},-Q_{1}Q_{2}^{*},
{1 \over \sqrt{2}}(Q_{1}Q_{1}^{*}-Q_{2}Q_{2}^{*}), Q_{2}Q_{1}^{*},  
{1 \over \sqrt{6}}(2Q_{3}Q_{3}^{*}-Q_{2}Q_{2}^{*}-Q_{1}Q_{1}^{*})}$,  \\
${-Q_{3}Q_{2}^{*},Q_{3}Q_{1}^{*}]}$,  ${[{\varepsilon}_{abc}Q_{a}Q_{b}Q_{c}Q_{m}]}$, 
$[Q_{a}^{*}Q_{m}]$ and  ${Q_{a}Q_{a}^{*}}$, (and their conjugates) respectively.   \\
Now in order to see the potential of superon-graviton model(SQM) as a composite model of matter  
we try to interpret the Feynman diagrams of SM(GUT) in terms of the superon pictures 
of all particles in  SM(GUT), i.e. a single line of a particle in the Feynman diagrams of SM(GUT) 
is replaced by multiple lines of superons  constituting the particle
with two  assumptoions at the vertex; (i) the analogue of the OZI-rule of the quark model and 
(ii) the superon number consevation. 
We find many remarkable results, e.g. in SM, naturalness of the mixing of $K^{0}$-$\overline{K^{0}}$, 
$D^{0}$-$\overline{D^{0}}$ and  $B^{0}$-$\overline{B^{0}}$, no CKM-like mixings among the lepton generations, 
${\nu_{e} \longleftrightarrow \nu_{\mu} \longleftrightarrow \nu_{\tau}}$ transitions beyond SM,  
strong CP-violation, small Yukawa couplings and no  $\mu \longrightarrow e + \gamma$ despite compositeness, etc. 
and in (SUSY)GUT, no dangerous diagrams for proton decay (without R-parity by hand), etc.\cite{ks1}\cite{ks2}. 
SGM may be the most economic one.      \par 
\section{Fundamental  Action  of  SGM }  
The supercharges $Q$ of Volkov-Akulov(V-A) model\cite{V-A} of the nonlinear SUSY(NL SUSY) is given 
by the supercurrents
\begin{equation}
J^{\mu}(x)={1 \over i}\sigma^{\mu}\psi(x)
-\kappa \{ \mbox{the higher order terms of $\kappa$, $\psi(x)$ and }\  \partial\psi(x) \}.
\end{equation}
(10) means the field-current identity between the elementary N-G spinor field
$\psi(x)$ and the supercurrent, which justifies our bold assumption
that the generator(super \\ 
charge) $Q^{N}$ (N=1,2,..10) of SO(10) SPA in the light-cone frame represents
the fundamental massless  particle, $superon$ $with$ $spin$ ${1 \over 2}$.
Therefore  the fundamental theory of SGM for spacetime and matter
at(above) the Planck scale is SO(10) NL SUSY in the
curved spacetime(corresponding  to the Clifford vacuum $\mid\Omega(\pm 2)\rangle$).
We extend the arguments of V-A to  high symmetric curved SGM spacetime, 
where NL SUSY SL(2C) degrees of freedom (i.e. the coset space coordinates representing N-G fermions) $\psi(x)$
in addition to  Lorentz SO(3,1) coordinates $x^{a}$ are embedded at every curved spacetime point with GL(4R) 
invariance. By  defining  a new tetrad   $ {w^{a}}_\mu(x)$,  $ {w_{a}}^\mu(x)$  and a new metric tensor 
$s^{\mu \nu}(x) \equiv {w_{a}}^{\mu}(x) w^{{a}{\nu}}(x)$  in SGM spacetime 
we obtain the following Einstein-Hilbert(E-H) type Lagrangian as the fudamental theory of SGM for  
spacetime and matter\cite{ks2}.
\begin{equation}
L=-{c^{3} \over 16{\pi}G}\vert w \vert(\Omega + \Lambda ),
\label{SGM}
\end{equation}
\begin{equation}
\vert w \vert=det{w^{a}}_{\mu}=det({e^{a}}_{\mu}+ {t^{a}}_{\mu}),  \quad
{t^{a}}_{\mu}={\kappa  \over 2i}\sum_{j=1}^{10}(\bar{\psi}^{j}\gamma^{a}
\partial_{\mu}{\psi}^{j}
- \partial_{\mu}{\bar{\psi}^{j}}\gamma^{a}{\psi}^{j}),
\label{w}
\end{equation} 
where $i=1,2, ..,10$,  $\kappa$ is a fundamental volume of four dimensional spacetime,  
${e^{a}}_{\mu}(x)$ is the vierbein of Einstein general relativity theory(EGRT) and 
$\Lambda$ is a  cosmological constant related to the superon-vacuum coupling constant. 
$\Omega$ is a new scalar curvature analogous to the Ricci scalar curvature $R$ of EGRT. 
The explicit expression of $\Omega$ is obtained  by just replacing ${e^{a}}_{\mu}(x)$  
by ${w^{a}}_{\mu}(x)$ in Ricci scalar $R$. 
The action  (\ref{SGM}) is invariant at least under  GL(4R), local Lorentz, global SO(10) 
and the following new (NL) SUSY transformation 
\begin{equation}
\delta \psi^{i}(x) = \zeta^{i} + i \kappa (\bar{\zeta}^{j}{\gamma}^{\rho}\psi^{j}(x)) \partial_{\rho}\psi^{i}(x),
\quad
\delta {e^{a}}_{\mu}(x) = i \kappa (\bar{\zeta}^{j}{\gamma}^{\rho}\psi^{j}(x))\partial_{[\rho} {e^{a}}_{\mu]}(x),
\label{newsusy}
\end{equation} 
where $\zeta^{i}, (i=1,..10)$ is a constant spinor and  $\partial_{[\rho} {e^{a}}_{\mu]}(x) = 
\partial_{\rho}{e^{a}}_{\mu}-\partial_{\mu}{e^{a}}_{\rho}$. 
These results can be understood intuitively by observing that 
${w^{a}}_{\mu}(x) ={e^{a}}_{\mu}(x)+ {t^{a}}_{\mu}(x)$  defined by 
$\omega^{a}={w^{a}}_{\mu}dx^{\mu}$, where $\omega^{a}$ is the NL SUSY invariant differential forms of 
V-A\cite{V-A}, and ${w^{a}}_{\mu}(x)$ and  $s^{\mu \nu}(x) \equiv {w_{a}}^{\mu}(x) w^{{a}{\nu}}(x)$ 
are formally  a new vierbein and a new metric tensor in SGM spacetime. 
In fact, it is not difficult to show the same behaviors of ${w_{a}}^{\mu}(x)$ and 
$s^{\mu \nu}(x)$ as those of ${e_{a}}^{\mu}(x)$ and $g^{\mu\nu}(x)$, i.e., 
${w_{a}}^{\mu}(x)$ and $s^{\mu \nu}(x)$ are invertible, 
${w_{a}}^{\mu} w_{{b}{\mu}} = \eta_{ab}$,  $s_{\mu \nu}{w_{a}}^{\mu} {w_{b}}^{\mu}= \eta_{ab}$, ..etc. 
and the following GL(4R) transformations of ${w^{a}}_{\mu}(x)$ and $s_{\mu\nu}(x)$ under (\ref{newsusy}) 
\begin{equation}
\delta_{\zeta} {w^{a}}_{\mu} = \xi^{\nu} \partial_{\nu}{w^{a}}_{\mu} + \partial_{\mu} \xi^{\nu} {w^{a}}_{\nu}, 
\quad
\delta_{\zeta} s_{\mu\nu} = \xi^{\kappa} \partial_{\kappa}s_{\mu\nu} +  
\partial_{\mu} \xi^{\kappa} s_{\kappa\nu} 
+ \partial_{\nu} \xi^{\kappa} s_{\mu\kappa}, 
\label{newgl4r}
\end{equation} 
where  $\xi^{\rho}=i \kappa (\bar{\zeta}^{j}{\gamma}^{\rho}\psi^{j}(x))$.
Therefore the similar arguments to EGRT in Riemann space can be carried out straightforwadly 
by using  $s^{\mu \nu}(x)$ (or ${w^{a}}_{\mu}(x)$) in stead of  $g^{\mu\nu}(x)$ (or ${e^{a}}_{\mu}(x)$), 
which leads to (\ref{SGM}) manifestly invariant at least under the above mentioned symmetries, 
which are isomorphic to SO(10) SP. 
The commutators of two new supersymmetry transformations  on $\psi(x)$ and  ${e^{a}}_{\mu}(x)$ 
are the general coordinate transformations 
\begin{equation}
[\delta_{\zeta_1}, \delta_{\zeta_2}] \psi
= \Xi^{\mu} \partial_{\mu} \psi,
\quad
[\delta_{\zeta_1}, \delta_{\zeta_2}] e{^a}_{\mu}
= \Xi^{\rho} \partial_{\rho} e{^a}_{\mu}
+ e{^a}_{\rho} \partial_{\mu} \Xi^{\rho},
\label{com1/2-e}
\end{equation}
where $\Xi^{\mu}$ is defined by
$\Xi^{\mu} = 2ia (\bar{\zeta}_2 \gamma^{\mu} \zeta_1)
      - \xi_1^{\rho} \xi_2^{\sigma} e{_a}^{\mu}
      (\partial_{[\rho} e{^a}_{\sigma]})$,
which form a closed algebra.                                                  \par
In addition, to embed simply the local Lorentz invariance
we follow EGRT formally and require that the new vierbein
$w{^a}_{\mu}(x)$ should also have formally
a local Lorentz transformation, i.e.,
\begin{equation}
\delta_L w{^a}_{\mu}
= \epsilon{^a}_b w{^b}_{\mu}
\label{Lrw}
\end{equation}
with the local Lorentz transformation parameter
$\epsilon_{ab}(x) = (1/2) \epsilon_{[ab]}(x)$.   
Interestingly,  we find that the following generalized 
local Lorentz transformations on  $\psi$ and $e{^a}_{\mu}$
\begin{equation}
\delta_L \psi(x) = - {i \over 2} \epsilon_{ab}
      \sigma^{ab} \psi,     \quad
\delta_L {e^{a}}_{\mu}(x) = \epsilon{^a}_b e{^b}_{\mu}
      + {\kappa \over 4} \varepsilon^{abcd}
      \bar{\psi} \gamma_5 \gamma_d \psi
      (\partial_{\mu} \epsilon_{bc})
\label{newlorentz}
\end{equation}
are compbtible with (\ref{Lrw}).  
[ Note that the equation (\ref{newlorentz}) reduces to the familiar
form of the Lorentz transformations
if the global transformations are considered, e.g., $\delta_{L}g_{\mu\nu}=0$.]
Also the local Lorentz transformation on $e{^a}_{\mu}(x)$ forms a closed algebra. 
\begin{equation}
[\delta_{L_{1}}, \delta_{L_{2}}] e{^a}_{\mu}
= \beta{^a}_b e{^b}_{\mu}
+ {\kappa \over 4} \varepsilon^{abcd} \bar{\psi}
\gamma_5 \gamma_d \psi
(\partial_{\mu} \beta_{bc}),
\label{comLr1/2}
\end{equation}
where $\beta_{ab}=-\beta_{ba}$ is defined by
$\beta_{ab} = \epsilon_{2ac}\epsilon{_1}{^c}_{b} -  \epsilon_{2bc}\epsilon{_1}{^c}_{a}$.
These arguments show that SGM action (\ref{SGM}) is invariant at least under\cite{st1}
\begin{equation}
[{\rm global\ NL\ SUSY}] \otimes [{\rm local\ GL(4,R)}] \otimes [{\rm local\ Lorentz}] 
\otimes [{\rm global\ SO(N)}].
\end{equation}
SGM  for spacetime and matter is the (isomorphic) case with N=10.   \par
\section{ Toward Low Energy Theory of SGM }     
For deriving the low energy behavior of 
the SGM action it is often useful to linearize  such a highly 
nonlinear theory and obtain a low energy effective theory 
which is renormalizable. 
Toward the linearization of the SGM we investigate the 
linearization of V-A model in detail.    %
The linearization of V-A model was investigated\cite{ik}\cite{r}  
and  proved that   N=1 V-A model of NL SUSY was equivalent to 
N=1 scalar supermultiplet action of L SUSY which was renormalizable.       
The general arguments on the constraints which gives the relations 
between the linear and the nonlinear realizations of global SUSY  
have been established\cite{ik}. 
Following the general arguments we  show  explicitly that nonrenormalizable 
N=1 V-A model is equivalent to a renormalizable total action of a U(1) gauge 
supermultiplet of the linear SUSY\cite{WZ} with 
the Fayet-Iliopoulos(F-I) $D$ term indicating a spontaneous SUSY breaking\cite{stt}. 
Remarkably we find that the magnitude of F-I $D$ term(vacuum value) 
is determined to reproduce the correct sign of  V-A action 
and that a U(1) gauge field  constructed explicitly in terms of N-G fermion fields 
is an axial vector for N=1.    \\
An $N=1$ U(1) gauge supermultiplet is given by a real superfield \cite{WB}
\begin{eqnarray}
V(x, \theta, \bar\theta) 
&=& C + i \theta \chi - i \bar\theta \bar\chi 
+ {1 \over 2} i \theta^2 (M+iN) 
- {1 \over 2} i \bar\theta^2 (M-iN) 
- \theta \sigma^m \bar\theta v_m        \nonumber  \\
& &+ i \theta^2 \bar\theta \left( \bar\lambda      
+ {1 \over 2} i \bar\sigma^m \partial_m \chi \right) 
- i \bar\theta^2 \theta \left( \lambda 
+ {1 \over 2} i \sigma^m \partial_m \bar\chi \right) 
 + {1 \over 2} \theta^2 \bar\theta^2 
\left( D + {1 \over 2} \Box C \right), 
\label{V}
\end{eqnarray}
where $C(x)$, $M(x)$, $N(x)$, $D(x)$ are real scalar fields, 
$\chi_\alpha(x)$, $\lambda_\alpha(x)$ and 
$\bar\chi_{\dot\alpha}(x)$, $\bar\lambda_{\dot\alpha}(x)$ are 
Weyl spinors and their complex conjugates, and $v_m(x)$ is 
a real vector field. 
We adopt the notations in ref.\ \cite{WB}. 
{}Following refs.\ \cite{ik},  we define the superfield 
$\tilde V(x, \theta, \bar\theta)$ by 
\begin{equation}
\tilde V(x, \theta, \bar\theta) = V(x', \theta', \bar\theta'), 
\label{tildev}
\end{equation}
\begin{equation}
x'^{\,m}  =  x^m + i \kappa \left( 
\zeta(x) \sigma^m \bar\theta 
- \theta \sigma^m \bar\zeta(x) \right), 
\theta'  =  \theta - \kappa \zeta(x), \qquad
\bar\theta' = \bar\theta - \kappa \bar\zeta(x). 
\label{cov}
\end{equation}
$\tilde V$ may be expanded as (\ref{V}) in component fields  
$ \{ \tilde\phi_i(x) \} =\{ \tilde C(x), \tilde\chi(x), \bar{\tilde\chi}(x), \cdots \}$, 
which can be expressed by $C, \chi, \bar\chi, \cdots$ and $\zeta$, $\bar\zeta$ 
by using the relation (\ref{tildev}). $\kappa$ is now defined with the dimension $(length)^{2}$. 
They have the supertransformations of the form 
\begin{equation}
\delta \tilde\phi_i = - i \kappa \left( \zeta \sigma^m \bar\epsilon 
- \epsilon \sigma^m \bar\zeta \right) \partial_m \tilde\phi_i. 
\end{equation}
Therefore, a condition $\tilde\phi_i(x) = {\rm constant}$ is 
invariant under supertransformations. 
As we are only interested in the sector 
which only depends on the N-G fields, 
we eliminate other degrees of freedom than the N-G fields 
by imposing  SUSY invariant constraints 
\begin{equation}
\tilde C = \tilde\chi = \tilde M = \tilde N = \tilde v_m 
= \tilde \lambda = 0, \qquad
\tilde D = {1 \over \kappa}. 
\label{constraints}
\end{equation}
Solving these constraints we find that the original component 
fields $C$, $\chi$, $\bar\chi$, $\cdots$ can be 
expressed by the N-G fields $\zeta$, $\bar\zeta$.
Among them, the leading terms in the expansion of the fields 
$v_m$, $\lambda$, $\bar\lambda$ and $D$, which contain 
gauge invariant degrees of freedom, in $\kappa$ are 
\begin{equation}
v_m  =  \kappa \zeta \sigma_m \bar\zeta + \cdot, 
\lambda  =  i \zeta 
- {1 \over 2} \kappa^2 \zeta 
\left( \zeta \sigma^m \partial_m \bar\zeta 
- \partial_m \zeta \sigma^m \bar\zeta \right) 
+ \cdot, 
%
%
D =  {1 \over \kappa} 
+ i \kappa \left( \zeta \sigma^m \partial_m \bar\zeta 
- \partial_m \zeta \sigma^m \bar\zeta \right) + \cdot, 
\label{relation2}
\end{equation}
where $\cdot$ are higher order terms in $\kappa$. 
Our discussion so far does not depend on a particular form 
of the action. We now consider a free action of a U(1) gauge 
supermultiplet of L SUSY with a Fayet-Iliopoulos $D$ term. 
In component fields we have 
\begin{equation}
S = \int d^4x \left[ -{1 \over 4} v_{mn} v^{mn} 
- i \lambda \sigma^m \partial_m \bar\lambda 
+ {1 \over 2} D^2 - {1 \over \kappa} D \right]. 
\label{gaugeaction}
\end{equation}
The last term proportional to $\kappa^{-1}$ is the 
Fayet-Iliopoulos $D$ term.
The field equation for $D$ gives 
$D = {1 \over \kappa} \not= 0$ in accordance with 
eq.\ (\ref{constraints}), 
which indicates the spontaneous breakdown of supersymmetry. 
We substitute eq.\ (\ref{relation2}) into the action 
(\ref{gaugeaction}) and obtain an action for the N-G fields 
$\zeta$, $\bar\zeta$ which is exactly N=1 V-A action. 
\begin{equation}
S = -{1 \over 2\kappa^2} \int d^4x \, \det \left[ \delta_m^n 
+ i \kappa^2 \left( \zeta \sigma^n \partial_m \bar\zeta 
- \partial_m \zeta \sigma^n \bar\zeta \right) \right]. 
\end{equation}
For N=1, U(1) gauge field  becomes 
$v_m \sim \kappa \bar\zeta \gamma_m \gamma_5 \zeta + \cdots$ 
in the four-component spinor notation, which is an axial vector. 
These are very  suggestive and favourable to SGM.           \par
\section{Discussion }
SGM action in SGM spacetime is a nontrivial generalization of E-H action in Riemann spacetime 
despite the liner relation 
${w^{a}}_{\mu} ={e^{a}}_{\mu}+ {t^{a}}_{\mu}$. 
In fact, by the redefinitions(variations) 
${e^{a}}_{\mu} \rightarrow {e^{a}}_{\mu}+ \delta {e^{a}}_{\mu}
={e^{a}}_{\mu}-{t^{a}}_{\mu}$ 
and  
$\delta {e_{a}}^{\mu}=-{e_{a}}^{\nu}{e_{b}}^{\mu}\delta {e^{b}}_{\nu}=+t{^{\mu}}_a$
the inverse 
${w_{a}}^{\mu}= e{_a}^{\mu}- t{^{\mu}}_a + t{^{\rho}}_a t{^{\mu}}_{\rho} 
- t{^{\rho}}_a t{^{\sigma}}_{\rho} t{^{\mu}}_{\sigma} 
+ t{^{\rho}}_a t{^{\sigma}}_{\rho} t{^{\kappa}}_{\sigma}t{^{\mu}}_{\kappa}$ 
does not reduce to ${e_{a}}^{\mu}$, i.e.  the  nonlinear terms in  
${t^{\mu}}_{a}$ in the inverse  ${w_{a}}^{\mu}$ can not be eliminated.
Because ${t^{a}}_{\mu}$ is not a metric. Such a redefinition on ${e^{a}}_{\mu}$ breaks the metric 
properties of ${w^{a}}_{\mu}$  and  ${w_{a}}^{\mu}$.  
Note that SGM action (\ref{SGM}) posesses two inequivalent flat spaces, 
i.e. SGM-flat ${w^{a}}_{\mu} \rightarrow {\delta^{a}}_{\mu}$ and 
Riemann-flat  ${e^{a}}_{\mu} \rightarrow {\delta^{a}}_{\mu}$. 
The expansion of (\ref{SGM}) in terms of ${e^{a}}_{\mu}$ and ${t^{a}}_{\mu}$ is a spontaneous breakdown 
of spacetime (\ref{SGM}) due to the degeneracy (\ref{newsusy}) from SGM to Riemann connecting 
with Riemann-flat spacetime\cite{st3}.
SGM (and V-A model with $N >1$)  posesses rich structures and the potential for defining 
completely the renormalizable (broken SUSY) models of the  local  field theory 
containing the massive high spin field by 
the lineralization in the curved spacetime. 
SGM for spin ${3 \over 2}$ N-G fermion\cite{st2} and SGM with the extra dimensions to be compactified 
are also in the same scope.
SGM cosmology is open.

\newpage

The author would like to thank  Alexander von Humboldt Foundation for the generous support and 
J. Wess for his encouragement, enlightening discussions and the warm hospitality  through the works. 
He is also grateful to M. Tsuda and Y. Tanii for the collaborations and K. Mizutani and  T. Shirafuji 
for useful discussions and  the hospitality at Physics Department of Saitama University.

\newpage

%
\newcommand{\NP}[1]{{\it Nucl.\ Phys.\ }{\bf #1}}
\newcommand{\PL}[1]{{\it Phys.\ Lett.\ }{\bf #1}}
\newcommand{\CMP}[1]{{\it Commun.\ Math.\ Phys.\ }{\bf #1}}
\newcommand{\MPL}[1]{{\it Mod.\ Phys.\ Lett.\ }{\bf #1}}
\newcommand{\IJMP}[1]{{\it Int.\ J. Mod.\ Phys.\ }{\bf #1}}
\newcommand{\PR}[1]{{\it Phys.\ Rev.\ }{\bf #1}}
\newcommand{\PRL}[1]{{\it Phys.\ Rev.\ Lett.\ }{\bf #1}}
\newcommand{\PTP}[1]{{\it Prog.\ Theor.\ Phys.\ }{\bf #1}}
\newcommand{\PTPS}[1]{{\it Prog.\ Theor.\ Phys.\ Suppl.\ }{\bf #1}}
\newcommand{\AP}[1]{{\it Ann.\ Phys.\ }{\bf #1}}

\end{document}